\documentclass[preprint2]{aastex}
\newcommand{\be}{\begin{equation}}
\newcommand{\ee}{\end{equation}}
\newcommand{\bea}{\begin{eqnarray}}
\newcommand{\eea}{\end{eqnarray}}
\begin{document}
\title{The Generation of Magnetic Fields Through Driven Turbulence}

\author{Jungyeon Cho \altaffilmark{1}}
\affil{Department of Astronomy, University of Texas, Austin, TX 
78712;cho@astro.as.utexas.edu}

\and

\author{Ethan T. Vishniac} 
\affil{Department of Physics and Astronomy, Johns Hopkins University,
Baltimore, MD 21218; ethan@pha.jhu.edu}
\altaffiltext{1}{Department of Physics and Astronomy, Johns Hopkins University,
Baltimore, MD 21218}


\begin{abstract}
We have tested the ability of driven turbulence to generate
magnetic field structure from a weak uniform field 
using three dimensional numerical simulations of incompressible
turbulence.  We used a pseudo-spectral code with  a numerical resolution of 
up to $144^3$ collocation points. We find that the magnetic 
fields are amplified through field line stretching at a rate 
proportional to the difference between the velocity and the 
magnetic field strength times a constant.
Equipartition between the kinetic and magnetic energy densities 
occurs at a scale somewhat smaller than the kinetic energy peak.
Above the equipartition scale the velocity structure is, as
expected, nearly isotropic. The magnetic field structure at these scales is
uncertain, but the field correlation function is very weak.
At the equipartition scale the magnetic fields show only 
a moderate degree
of anisotropy, so that the typical radius of curvature of field lines is
comparable to the typical perpendicular scale for field reversal. 
In other words, there are few field reversals within eddies at the
equipartition scale, and no fine-grained series of reversals at
smaller scales.  At scales below the equipartition
scale, both velocity and magnetic 
structures are anisotropic; the eddies are stretched along the local 
magnetic field lines, and the magnetic energy dominates the kinetic
energy on the same scale by a factor which increases at higher
wavenumbers.  We do not show
a scale-free inertial range, but the power spectra are a function
of resolution and/or the imposed viscosity and resistivity.  Our
results are consistent with the emergence of a scale-free inertial
range at higher Reynolds numbers.

\end{abstract}

\section{INTRODUCTION}
There are at least two distinct types of magnetohydrodynamic (MHD) turbulence.
When the external large scale magnetic field is strong, the resulting turbulence
can be described as the nonlinear interactions of Alfv\'en waves 
(e.g., Goldreich and Sridhar 1995, 1997).
In contrast, when the external field is weak, MHD turbulence
near the scale of the largest energy containing eddies
will be more or less like ordinary hydrodynamic turbulence with a small 
magnetic back 
reaction. In this regime, the turbulent eddy turnover time 
at the large scale ($L/V$) is less than
the Alfv\'enic time of the scale ($L/B$), where $V$ and $B$ are r.m.s. velocity
and magnetic field strength (divided by $(4\pi\rho)^{1/2}$) 
respectively and $L$ is the scale of energy
injection or the largest energy containing eddies.

Various aspects of weak/zero external field MHD turbulence have been studied
both theoretically and numerically. One of the most important issues in this 
regime is the generation of large scale fields.
Since large scale magnetic fields are observed in almost all astrophysical
objects, the generation and maintenance of such fields is of great importance.
In mean field dynamo theory (see Moffatt 1978; Parker 1979),
turbulent motions at small scales are biased to create a 
non-zero electromotive force along the direction of the large-scale magnetic field.
This effect (the `$\alpha$-effect') works to amplify and maintain
large-scale magnetic fields.  Whether or not this effect actually
works depends on the structure of the MHD turbulence, 
especially on the mobility of the field lines.  
For example,  Vainshtein and Cattaneo (1992)
have argued that when equipartition between magnetic and kinetic
energy densities occurs at any scale larger than the dissipation scale, 
the mobility of the field lines and the $\alpha$ effect will be greatly 
reduced.  An alternative approach, which may avoid such difficulties, 
is to appeal to small scale field
line stretching alone to generate strong magnetic fields (Batchelor 1950,
Kazantsev 1967).  In this model large scale magnetic fields are actually
composed of parallel fibrils of random polarity, i.e. the large scale
field is defined by the preferred axis of the magnetic field direction,
but there is little large scale flux in such fields.  Whether or not
the mean field dynamo mechanism works, it is interesting to contemplate
whether or not such fine-grained fields can be produced in the absence
of any $\alpha$ effect.

In addition, there are at least two other points of contention regarding
the nature of magnetic fields generated in MHD turbulence.  First,
the mobility of magnetic fields will be affected by their intermittency.
Various models for the generation and dynamics of flux tubes have
been proposed, motivated in part by observations of the solar
photosphere (e.g., Vishniac 1995a, 1995b; Brandenburg at al. 1995).
However, so far numerical simulations have tended to show only modest
levels of intermittency.  Second, there is the nature of the
energy cascade and power spectrum of MHD turbulence.  
There are numerous analytical models for dealing with 
MHD turbulence: the eddy-damped
quasinormal Markovian (EDQNM) approximation (Pouquet et al. 1976), 
the renormalization group technique (Fournier at al. 1982), and others,
including recent work by Goldreich and Sridhar (1995, 1997) in which
they treated nonlinear eddy interactions as a series of interactions
between marginally nonlinear Alfv\'en waves.  The last is perhaps
the most promising.
The energy spectra of the MHD turbulence is one of the most poorly understood
fields in astrophysics.
Early work by Iroshnikov (1963) and Kraichnan (1965) obtained a $k^{-3/2}$ 
spectrum for both
magnetic energy and kinetic energy in the presence of a dynamically
significant magnetic field.  However, this work was based on assumption of 
isotropy in wavenumber space, which is difficult to justify unless the
magnetic field is very weak.
Goldreich and Sridhar's model assumes a critical level of anisotropy,
such that magnetic and hydrodynamic forces are comparable, and
predicts a $k^{-5/3}$ spectrum  for strong external field turbulence
(which should apply to very small scales within any MHD turbulent cascade).  
Solar wind observations, which
are well within the strongly magnetized regime, show an energy
spectrum $E(k) \sim k^{-1.70}$ (Leamon et al. 1998).  Numerical
studies have only recently been able to address this question.
Recent work by Cho and Vishniac (2000) and Maron and Goldreich (1999)
seems to support the scaling laws of Goldreich and Sridhar.

Here we will concentrate on the structure of magnetic fields
in MHD turbulence when the external field is weak.
In the case of hydrodynamical turbulence, the energy cascades to smaller 
scales.  If we introduce a weak uniform magnetic field, turbulent motions will
stretch the magnetic field lines and divert energy to the
small scale magnetic field.
As the magnetic field lines are stretched, the magnetic
energy density increases rapidly, until
the generation of small-scale magnetic structures 
is balanced by the magnetic back reaction at some scale between 
$L$ and the dissipation scale.
This will happen when the magnetic and kinetic energy densities
associated with a scale $l$ are comparable so that Lorentz forces 
resist further stretching at or below that scale\footnote{The first 
energy equipartition scale $l$ can be larger than 
the dissipation scale when the equipartition occurs before the energy
cascade reaches the dissipation scale.}. 
However, 
stretching at scales larger than $l$ is still possible, and
the magnetic energy density will continue to grow if $l$ can
increase\footnote{Our results show that $l$ does increase and the 
magnetic energy
continues to grow.}.
Eventually, a final stationary state will be reached.

This discussion raises two questions. What is the scale of
energy energy equipartition? What is the magnetic 
field structure on this scale?
The answer to the latter question will depend on the nature of diffusive 
processes acting on the magnetic field.
First, suppose that magnetic field lines are unable to smooth the tangled 
fields at small scales. Then, as a result of the turbulent energy cascade and
the subsequent stretching of magnetic field lines, magnetic fields have
thin fibril structures with many polarity reversals within the 
energy equipartition scale $l$.  Consequently,
magnetic structures on the equipartition scale are highly
elongated along the local magnetic field direction.
This is the kind of picture one obtains by considering passive
advection of magnetic fields in a chaotic flow (for a review see Ott 1998).
The degree to which this can be applied to a realistic, highly conducting
fluid is controversial.
On the other hand, if we assume MHD turbulence is always capable 
of relaxing tangled field lines at small scales, then we expect 
eddies at the 
final equipartition scale to be nearly isotropic. 
That is, the typical radius of curvature of field lines will be
comparable to the scale of field reversal.  
This picture does not imply a clear expectation for
the scale of equipartition in the stationary
state. As we will see later, our simulations support the latter picture
but with the energy equipartition
scale near, but somewhat smaller than, the largest energy containing 
eddy scale.

In this paper, we will interpret the results of numerical simulations
of MHD turbulence with a weak external field in terms of these
models.
In \S2, we discuss the numerical method we used and performance of the code.
In \S3, we present the results of simulations.  Finally,
\S4 contains a discussion of the results and our conclusions.

\section{NUMERICAL METHOD AND PERFORMANCE OF THE CODE}
We used a pseudo-spectral code to solve the MHD equations in a periodic box of size
$2\pi$:
\begin{equation}
\frac{\partial {\bf V} }{\partial t} = (\nabla \times {\bf V}) \times {\bf V}
      - (\nabla \times {\bf B})
        \times {\bf B} + \nu \nabla^{2} {\bf V} + {\bf f} + \nabla P' ,
        \label{veq}
\end{equation}
\begin{equation}
\frac{\partial {\bf B}}{\partial t}= {\bf B} \cdot \nabla {\bf V}
     - {\bf V} \cdot \nabla {\bf B} + \eta \nabla^{2} {\bf B} ,
     \label{beq}
\end{equation}
\be
      \nabla \cdot {\bf V} =\nabla \cdot {\bf B}= 0,
\ee
where $\bf{f}$ is a random forcing term with unit correlation time,
$P'\equiv P + {\bf V}\cdot {\bf V}/2$, 
${\bf V}$ is the velocity, and ${\bf B}$ is magnetic field divided by
$(4\pi \rho)^{1/2}$. Thus the field ${\bf B}$ is, in fact,
the Alfv\'{e}nic velocity. The Alfv\'en velocity of
the background field, $B_0$,
is set to $10^{-3.5}$ in all simulations.
Throughout the paper, we consider only cases where
viscosity is equal to magnetic diffusivity:
\be
  \nu = \eta.
\ee
In the pseudo-spectral approach, the actual calculations are 
performed in wavevector space. 
The nonlinear terms are evaluated in real space using 
Fourier transformed variables and their derivatives and then
transformed back into their Fourier components.
The forcing term consists of 21 Fourier components 
with $2\leq k \leq \sqrt{12}$. 
The peak of energy injection is at $k\approx 2.5 (\equiv k_L)$.
We adjusted the amplitudes of the forcing components so that $V \approx 1$.
We use exactly the same forcing terms for all simulations.
Therefore, the external magnetic fields are very weak for all simulations
(i.e. $B_0 \ll V$).
Each forcing component consists of two parts: a linearly 
polarized component and a small circularly polarized 
component with a preferred helicity.
The latter was introduced to provide non-zero helicity injection to
the turbulence. The resulting fluid helicity is $0.3\sim 0.4$ for 
all runs.
We use an appropriate projection operator to calculate the 
$\nabla P'$ term in
Fourier space and also to enforce the divergence-free condition
($\nabla \cdot {\bf V} =\nabla \cdot {\bf B}= 0$).
We use up to $144^3$ collocation points.
We use the integration factor technique for kinetic and 
magnetic dissipation terms
and the leap-frog method for the nonlinear terms.
At $t=0$, the magnetic field is uniform and the velocity is spread
between $2\leq k \leq 4$ in wavevector space.
We give the parameters for each run in Table 1.

Either physical viscosity (and diffusivity) or 
hyperviscosity (and hyperdiffusivity)
is used in the dissipation terms (see Table 1).
The power of hyperviscosity
is set to 8, such that the dissipation term in the above equation
is replaced with
\be
 -\nu_8 (\nabla^2)^8 {\bf V},
\ee
where $\nu_8$ is determined from the condition $\nu_h (N/2)^{2h} \Delta t
\approx 0.5$ (see Borue and Orszag 1995,1996). 
Here $\Delta t$ is the time step and $N$ is the number
of grids in each direction.
The same expression is used for the magnetic dissipation term.

        When ${\bf f}=\nu=\eta=0$, that is, without forcing or dissipation,
        the total energy drops by 0.87 percent after 13 time units,
        apparently due to truncation errors.
        In the dissipative case the injection and dissipation of energy are well
        balanced after the magnetic energy reaches the stationary state:
        \be
        <\epsilon> = <D_K> + <D_M>,
        \ee
        where $\epsilon$ is the energy injection rate and 
        $D_K$ and $D_M$ are the kinetic and the magnetic energy dissipation 
        rates respectively. 
        The angle bracket stands for an appropriate space-time average,
        which is taken after the magnetic energy reaches the stationary state.
        
Before we describe the third test, we define several transfer
functions.
Imagine velocity components in a thin (i.e., thickness=1) 
spherical shell 
of radius $k$ in Fourier space.
The work done to the velocity components by magnetic fields can be 
written as follows:
\be
  -\sum_{k\leq |{\bf p}| < k+1} \hat{\bf V}( {\bf p})^*\cdot 
  \widehat{\left[(\nabla \times {\bf B})\times {\bf B} \right]}( {\bf p}),
\ee 
where {\it hatted} variables are Fourier space variables and `*' 
denotes a complex conjugate.
This relation is derived from the second term on the right-hand side of 
equation (1).
This quantity is the $negative$ of the energy transferred 
to the magnetic fields from the velocity components per unit time:
\be
 T_{b|v_k}(k)= +\sum_{k\leq |{\bf p}| < k+1} \hat{\bf V}( {\bf p})^*\cdot 
  \widehat{\left[(\nabla \times {\bf B})\times {\bf B} \right]}( {\bf p})..
\ee
Note that the subscripts of $T_{X|Y}$ read `{\it to X from Y}.'
Similarly, from the first term in the right-hand side of
equation (1), we obtain $T_{v|v_k}(k)$:
\be
  T_{v|v_k}(k) = 
  -\sum_{k\leq |{\bf p}| < k+1} \hat{\bf V}( {\bf p})^*\cdot 
  \widehat{\left[(\nabla \times {\bf V})\times {\bf V} \right]}( {\bf p}).
\ee
Now, consider the magnetic components in the same shell.
{}From the first term in the right-hand side of equation (2),
we obtain the energy transferred 
to the velocity fields from the magnetic components per unit time:
\be
  T_{v|b_k}(k) = 
  -\sum_{k\leq |{\bf p}| < k+1} \hat{\bf B}( {\bf p})^*\cdot 
  \widehat{\left[{\bf B} \cdot \nabla {\bf V} \right]}( {\bf p}).
\ee
the work done to the magnetic components by velocity fields:
Again, this quantity is the $negative$ of the work done to the magnetic 
components by velocity fields per unit time.

In our third test, we checked the energy budget 
of the magnetic fields. 
The $net$ energy transferred to magnetic fields from velocity fields can be 
calculated from either $T_{b|v_k}(k)$ or $T_{v|b_k}(k)$.
This energy will
disappear through diffusive damping.
Therefore, the numerical scheme should satisfy 
\be 
  T_{b|v} = -T_{v|b} = D_M,
\label{balance}
\ee
where
\begin{eqnarray}
  T_{b|v} \equiv \sum_{k=0}^{k_{max}} T_{b|v_k}(k), \\
  T_{v|b} \equiv \sum_{k=0}^{k_{max}} T_{v|b_k}(k), 
\end{eqnarray}
and $k_{max}$ is the largest wave number in Fourier space and is
equal to $N/2$.
{}Fig. 1 shows that our code satisfies equation (\ref{balance}).
The two curves $T_{b|v}(t)$ and $-T_{v|b}(t)$ coincide exactly.
It is interesting to note that there is a time delay between
$T_{b|v}$ ($=T_{v|b}$) and $D_M$, which is a measure of
the energy cascade time.

\section{RESULTS}
\subsection{Generation of Magnetic Fields}
We list the results of our simulations in Table 1.
We obtained $V^2$, $B^2$, $\epsilon$, and $D_M$ by averaging over 
$(t_1, t_2)$ using all available data. 
However, $T_{v|v_2}$ ($\equiv T_{v|v_k}(k=2)$) 
and $T_{b|v_2}$ ($\equiv T_{b|v_k}(k=2)$) were calculated from a sparse 
subset of the data.
It is important to note that, unless otherwise noted, 
these time averages are taken after
the turbulence has reached the stationary state.
Conclusions based on these averages do not apply to the initial
growth phase of the magnetic field.

{}Fig. 2 shows time evolution of kinetic and magnetic energy density.
All the simulations have similar kinetic energy densities.
However, the magnetic energy density obviously depends on the
ohmic diffusivity $\eta$.
After an initial growth phase, the magnetic energy reaches a
stationary state.
The initial growth rate of the magnetic energy depends on diffusivity.
Even though it is not clearly shown in the figure, the initial growth
phase consists of two stages for Run 144A, the hyperviscosity run.
In the first stage, which begins at t=0 and
ends after a few dynamical times, the growth rate is very fast.
At this stage the
magnetic energy grows through the stretching of 
magnetic field lines.  Growth is fast because there is
no significant back reaction by the magnetic fields.
{}Fig. 3 shows that
stretching is initially most active near, or
somewhat larger than, the dissipation scale (cut-off scale) and
the magnetic energy spectrum peaks at this scale.
As the magnetic energy grows, the 
magnetic back reaction becomes important
at the dissipation scale. 
When energy equipartition is reached at this scale,
the stretching rate slows down and a second stage of slower
growth begins.  Fig. 3 shows that during this stage
the peak of the magnetic power spectrum moves to larger scales.  
{}Fig. 2 shows the second stage ends at $t=30\sim 40$ for Run 144A. 
Other runs with physical viscosities (Run 128A, 96A, 64A, etc.)
also show similar behavior.
However, the detailed evolution seems different from the 
hyperviscosity case.
{}For example, 
the diffusivity also affects
the net growth rate in the second stage. 

In Fig. 4a, we plot the kinetic and magnetic energy densities as 
functions of $\nu$ $(=\eta)$.  Error bars are estimated from
\be
  2 \frac{ < X^2 > - <X>^2 }{ \sqrt{ (t_2-t_1)/t_{corr} } },
\ee
where $<\cdot>$ denotes the time average and $X=V$ and $B$.
As noted above, this
average is taken after the turbulence has reached its stationary
state.
The correlation time $t_{corr}\sim 3$ is  calculated directly.
We see that the kinetic energy is almost independent of 
$\nu$ $(=\eta)$.  
The constancy of the kinetic energy density is the result of
       the magnetic field. In purely hydrodynamic simulations
the kinetic energy density increases as $\nu$ decreases 
       (see REF2 and REF3 in Table 1).
On the other hand, the magnetic energy increases 
as $\nu$ $(=\eta)$ decreases.
This figure suggests that at small $\eta$ even a weak diffuse magnetic 
field can lead to a strong final state. 
In the case of hyperviscosity, the magnetic energy is 
almost half of the kinetic energy at late times.
This is the result of field line stretching, and 
        not the alpha effect (see next paragraph and \S 4).
The energy injection rates 
$\epsilon$ ($\equiv {\bf f}\cdot {\bf V} = D_K+D_M$)
are independent of $\nu$ $(=\eta)$ and in the 
range $0.161< \epsilon<0.166$ (Table 1).

In Fig. 4b, we plot an empirical relation for 
the generation of magnetic fields in the stationary state:
\be
  D_M \sim \left( \frac{V}{L} - c \frac{B}{L} \right) B^2,
  \label{stretching}
\ee
where $c$ is a constant ($\sim 0.63$).
It is a somewhat dangerous to assume that Run 144A, the
        hyperviscosity run, is equivalent to a the case of vanishing
        viscosity.  In particular, Fig. 5 shows a sharp change in the
        energy transfer to the magnetic field near the dissipation
        scale.  This makes it difficult to extend this relationship
        to the case of vanishing diffusivity.\footnote{In Run 144A, 
        the magnetic dissipation is given by $\sum \nu_8 k^{16} |\hat{\bf B}|^2$.
        Although we can calculate this quantity directly,
        we should not use this for equation (15) because
        $D_M$ in equation (15) actually represents
        the amount of energy transferred to the magnetic field on large
        scales.  If we use a physical viscosity (and diffusivity), the two
        quantities are same.
        However, in hyperviscosity runs, the two quantities diverge
        because nonlinear processes
        occurring near the dissipation cut-off are presumably unphysical.
        The dotted curve in Fig. 5 represents the amount of
        energy transferred to the magnetic field in Run 144A.
        The curve shows a non-negligible change after the dissipation
        cut-off ($k_d \sim 50$), which means that nonlinear processes
        do occur within the dissipation range (i.e. $k>k_d$).
        If we take the value of $\Pi_{b|v_k}$ at $k\sim k_d$ for
        $D_M$ and substitute it into equation (15), then we obtain
        $(v-B/1.6)B^2/D_M = 1.24$, in good agreement
        with the other data shown in Fig. 4b.
        This suggests that equation (15) is is true even in the limit
        of vanishing viscosity (and diffusivity), but given the
        existence of anomalous energy transfer near the dissipation scale,
        this conclusion may be premature.}.

What is the origin of this correlation?
In the stationary state,
the magnetic dissipation ($D_M$) is balanced by the net energy
transferred to the magnetic fields from the velocity fields.
The right-hand side therefore tells us that the net energy
transferred to the magnetic fields is proportional to large-scale eddy
turnover rate ($V/L$) minus an 
Alfv\'{e}nic frequency ($B/L$) times a constant.
The large-scale eddy turnover rate is equal to the stretching rate
of the magnetic field when the back reaction is zero.
We identify the second term on the right-hand side of this 
equation as the effect of the magnetic back reaction.

This interpretation is supported by the spectrum of the 
energy transfer rate, $T_{b|v_k}(k)$. 
The function $T_{b|v_k}(k)$ is the energy lost by 
velocity components within
a unit shell of radius $k$ in wavevector space 
through interactions with the magnetic field.
When this quantity is positive,
it means that energy is being transferred from 
velocity components in the unit shell to the magnetic field.
In Fig. 5a, we plot $ T_{v|v_k}(k)$, $ T_{b|v_k}(k)$ and $ T_{v|b_k}(k)$
for Run 144A.
{}From the spectrum of 
$T_{b|v_k}(k)$ we can see that magnetic fields are
driven by energy extracted from the large scale velocity components
(cf. Kida et al., 1993).
We can explain this result if we suppose that the magnetic field 
energy is generated by field line stretching  at large scales.
On the other hand, the magnetic fields have a small rate of
net energy transfer to small scale motions, 
implying that on small scales the turbulent motions can be
described as nonlinear Alfv\'en waves, with a rough balance
between kinetic and magnetic energy densities.

In Fig. 5b, we plot the spectra of energy fluxes.
The energy flux $\Pi_{b|v_k}(k)$ is the amount of energy transferred
from velocity components whose wavenumbers are less than or equal to $k$
to the magnetic components of all wavenumbers:
\be
  \Pi_{b|v_k}(k) = \int_0^k T_{b|v_p}(p) dp.
\ee
We define other fluxes similarly.
The graph of $\Pi_{b|v_k}(k)$ rises rapidly at large scales ($k<10$) and
reaches the value of energy injection rate $\epsilon$ ($\sim 0.16$) at
$k\sim 10$. 
This means that the energy injected to drive the turbulence 
($\sim 0.16$ per unit time) 
is absorbed at large scales by the magnetic fields. 
On smaller scales ($k>10$) the velocity field does not drive
the magnetic fields, but instead  $absorbs$ energy.  In summary,
the rise of $\Pi_{b|v_k}(k)$ at small wavenumbers is due to field
line stretching and
the decrease at large wave numbers is a sign of 
Alfv\'{e}nic turbulence at small scales. 
Consequently, at small scales, we can treat the turbulent
motions as fully developed MHD turbulence, even when external
fields are weak.

Before proceeding, it is useful to take note of a property of 
hyperviscosity simulations.
The quantity $\Pi_{v|v_k}(k) + \Pi_{b|v_k}(k)$ is the total energy
transferred from large scale velocity fields ($|{\bf k}| < k$)
to either magnetic fields or small scale velocity fields. 
When there is no dissipation, 
the energy injected by the driving force is either transferred
to magnetic fields or transferred to small scale velocity fields: 
\be
 \epsilon = \Pi_{v|v_k}(k) + \Pi_{b|v_k}(k).
\ee
That is, the quantity $\Pi_{v|v_k}(k) + \Pi_{b|v_k}(k)$
has to be a constant for a hyperviscosity simulation.
In fact, in the hyperviscosity simulation (Run 144A), 
this quantity is constant for $k<50$.

In Fig. 6, we plot $T_{v|v_2}$ and $T_{b|v_2}$.
Here $ T_{v|v_2}$ ($\equiv T_{v|v_k}(k=2)$) 
is proportional to the rate at which the energy of large scale eddies
is transferred to small scale velocity fields. 
On the other hand, 
$ T_{b|v_2}$ ($\equiv T_{b|v_k}(k=2)$) is proportional to  
the rate at which energy of large scale eddies
is transferred to magnetic fields.
Note that $k=2$ corresponds to the scale of largest energy containing
eddies.
Fig. 6a shows
\be
  T_{v|v_2} \sim V^3,
\ee
which suggests that 
large scale kinetic energy ($V^2$) is transferred to smaller
scales within an eddy turnover time ($L/V$).
That is, turbulence at large scale is broadly similar to 
ordinary hydrodynamic turbulence 
and the classical energy cascade model seems to work.
In the case of hyperviscosity, the figure suggests that
        the energy transfer rate
        may not exactly follow the scaling relation given above.
        However, as we can see in the figure, 
        the deviation, if any, will be very small.

{}Fig. 6b shows
\be
  T_{b|v_2} \sim B^2 V.
\ee
When magnetic field is generated through stretching of field lines, 
$ T_{b|v_2}$ will be proportional to the stretching rate at large scales.
Although equation (\ref{stretching}) implies that overall stretching rate 
is proportional to $(V-const\cdot B)$, 
{}Fig. 6b suggests that the stretching rate scales as $B^2 V$
at the largest energy containing eddy scale.
That is, it is proportional to the large eddy turnover rate $V/L$.

\subsection{The Structure of Turbulence}

We plot energy spectra in Fig. 7.
{}For all simulations using physical viscosities (runs 128A, 96A, 
72A, and 64A-D) the
kinetic energy spectra are almost independent of $\nu$ at large 
scales ($k<10$).
However, this is not true in the case of hyperviscosity (Run 144A).
Magnetic energy spectra peak near, but somewhat smaller than, 
the energy injection scale.
The position of the peaks depends on magnetic diffusivity.
We see that the location of the peaks
moves towards the largest scale as diffusivity increases.
In the case of hyperviscosity (and hyperdiffusivity) simulation,
both kinetic and magnetic spectra show the well known bottleneck 
effect (see, for example, Borue and Orszag 1996) for $k>\sim 20$.
This effect is characterized by a local enhancement of the 
energy spectra at scales somewhat larger than the dissipation cutoff.

We define the energy equipartition wavenumber $k_e$ such that
\be
  \int_0^{k_e} dp E_M(p) = \int_{k_e}^{k_{max}}dp E_K(p).
\ee
In this definition, energy equipartition between 
$large$ scale magnetic fields
and $small$ scale velocity fields occurs at $k=k_e$.
(Therefore, $k_e$ is different from the scale of equipartition
between $small$ scale magnetic energy and $small$ scale kinetic energy.
A good measure for the equipartition scale between small scale energies 
is
the wavenumber at which $E_K(k)=E_M(k)$. In the case of hyperviscosity,
the two wavelengths are similar.)
{}From Table 1, we can see that $k_e$ is not very sensitive to the value
of viscosity.
In all cases, the equipartition scale ($\sim 1/k_e$) is somewhat smaller 
than the peak of the kinetic energy spectrum ($1/k_L \sim 1/2.5$).
The magnetic fields associated wavenumbers greater than $k_e$ 
act like a strong uniform external magnetic field for eddies 
smaller than $\sim 1/k_e$, and the structure of turbulence at
smaller scales will be similar to Alfv\'{e}nic turbulence.
On the contrary, turbulence at scales larger than $\sim 1/k_e$
will be more or less like ordinary hydrodynamic turbulence.
This argument is supported by the fact that
the value of $k_e$ is very close to the peak of 
$\Pi_{b|v_k}(k)$ for all runs.
For example, Fig. 5b shows $\Pi_{b|v_k}(k)$ peaks at $k \sim 10$ for 
Run 144A.
Table 1 shows that $k_e \sim 8.4$ for the same run.
(Our analysis also shows that, 
for Run 96A, $k_e \sim 6.7$ and $\Pi_{b|v_k}(k)$ peaks at $k \sim 8$.
And, for Run 64A, $k_e \sim 6.6$ and $\Pi_{b|v_k}(k)$ peaks at $k \sim 8$...)
This suggests that, below the equipartition scale, 
       the turbulent fluid is in a state of nonlinear Alfv\'enic
       turbulence, with a fixed ratio between the kinetic and magnetic 
energies associated with perturbations on these scales.

We see that a small external field leads to local magnetic energy
density which, in the limit of very small diffusivity, 
is comparable to the kinetic energy density, with
a typical scale only slightly smaller than the scale at which
the turbulence is driven.  This still leaves the question of
whether the magnetic field is characterized by numerous polarity
reversals within each eddy.  We have examined this question
by considering the magnetic field second order structure function.
A detailed description of our method can be found in Cho
and Vishniac (1999).  Here we note only that we define the
structure function in cylindrical coordinates as
\be
F(\rho,z) =\langle<|{\bf Y}({\bf r}_1)-{\bf Y}({\bf r}_2)|^2\rangle,
\ee
where ${\bf Y}$ can be either the magnetic field or the velocity.
The coordinates are defined relative to the
{\it local} field direction, with $\rho$ being the distance 
perpendicular to $({\bf B}({\bf r}_1)+{\bf B}({\bf r}_2))/2$
and $z$ the distance parallel to it.

We plot the results for run 144A in Fig. 8.  
The shape of the contours represents the average shape of the eddies, which 
in turn reflects the degree of anisotropy along the local field direction.
As expected, the 
velocity field shows isotropy at large scales and anisotropy
at small scales. The magnetic fields show an insignificant
amount of correlation at large scales and strong anisotropy
at small scales.  At intermediate scales, comparable to the
equipartition scale, the magnetic field shows a factor of
two difference in correlation lengths along local 
field line directions and across them.  The implication is that
the field is {\it not} composed of numerous parallel fibrils with
frequent polarity reversals on small scales, as envisioned by
Batchelor (1950) and Kazantsev (1967).  
Instead, the field is largely smoothed by small scale turbulent 
diffusion, and shows only one or two polarity reversals at
its typical scale of organization.  

\section{DISCUSSION AND CONCLUSIONS}

We have presented results of MHD numerical simulations for unit magnetic
Prandtl number (i.e. $\nu=\eta$). A weak diffuse magnetic field was 
used as an initial condition, and we tested the sensitivity of
our results to dissipative effects by varying the viscosity and
numerical resolution in our simulations.
We have found that magnetic fields are amplified through field line
stretching at a rate proportional to $(V-cB)/L$, where $c\sim 0.63$.
It is not certain whether or not the hyperviscosity run follows
this relation.
We have also shown that, in the limit of $\nu$ $(=\eta) \rightarrow 0$, 
the magnetic field reaches a final stationary state where the
magnetic energy density is comparable to the kinetic energy.
Since the simulations using physically realistic functional forms 
for the diffusivity are very far away from this limit, we are
not able to extrapolate to a definite final ratio of $E_B/E_K$,
but it should be at least as large as the value reached in
run 144A, $\sim 0.6$.
The typical scale of the magnetic field is slightly smaller
than the scale of the largest energy containing eddies.
At or above the equipartition scale, eddies are almost isotropic.
However, eddies smaller than the equipartition scale show elongation 
along the local magnetic field lines.

When we vary viscosity (and diffusivity) explicitly,
physical quantities scale as follows:
\begin{enumerate}
  \item  $V^2 \sim$ const
  \item $D_M \sim (V-c B)B^2$, c=constant
  \item $T_{v|v_k}(k=2) \sim V^3$
  \item $T_{b|v_k}(k=2) \sim VB^2$.
\end{enumerate}

The turbulent dynamo effect did not play a role in these
simulations.  There is no sign of the spontaneous generation
of a large scale field, above what we would expect by extrapolating
the magnetic energy power spectrum to wavenumbers smaller than the
energy peak.  Our results are most plausibly explained in terms
of random field line stretching within eddies.  We have tested this
conclusion by conducting additional simulations with no imposed helicity.
The results were almost identical to the simulations presented here.  
However, 
this does not allow us to draw any general conclusions about the
possibility of mean-field dynamos in astrophysical objects.  We
note that the average helicity for all the runs listed in Table 1 is 
in the range $0.3$ to $0.4$.  Since the smallest allowable wavenumber 
is $1$, this implies a maximum mean-field growth rate of order
$0.1$ times the eddy correlation time.  By comparison, the 
turbulent dissipation rate is $\langle V^2\rangle/3$ (or $\sim 0.3$) 
times the eddy correlation time.  The obvious conclusion is that
no dynamo effect is expected.   The energy injection scale here
is too close to the size of the computational box. 

The isotropic structure of eddies at large scales in our
simulations implies something about diffusive processes in MHD turbulence.
In particular, it suggests the presence of an effective magnetic
diffusivity which is tied to the large scale eddy size rather than
the resistive scale.  This is an old idea (see, for example, Parker 1955), 
and is equivalent
to the notion that the usual turbulent diffusion coefficients can
be substituted for ohmic resistivity.  In its original form
it was based on a picture of small scale field line mixing, which
has been convincingly criticized by Parker (1992).  However, it
can also be justified by appealing to rapid reconnection, that is,
field line reconnection at rates which are comparable to an eddy
turn over rate and  much more rapid
than estimates based on the Sweet-Parker reconnection rate (Sweet 1958, 
Parker 1957). In terms of our simulations, we note that if reconnection is 
slow than we expect that magnetic fields will
show a thin fibril structure in the saturated state, with a typical 
field reversal scale much smaller than the curvature
scale. We see from the structure function that this is not the case.
Apparently, in our simulations at least, reconnection is fast enough to 
relax tangled magnetic field structures in no more than an eddy turn over
time.  Either the suppression of reconnection appears suddenly at 
higher resolutions, or theoretical arguments suggesting fast reconnection
in highly conducting turbulent fluids (Lazarian and Vishniac 1999) are
correct.  If the latter interpretation holds up, then 
arguments suggesting the suppression of the $\alpha$ effect
(e.g. Vainshtein and Cattaneo 1992) at high Reynolds numbers are unlikely to 
prove correct.

In what follows, we will give a rough explanation for the 
dependence of field quantities on $\nu$ (=$\eta$).
Readers are advised that this analysis might fail
    in the limit of vanishing diffusivity. The reason is that
    run 144A shows slight deviations from the extrapolation results
    (see $V^2$ in Fig. 4a and $T_{b|v_2}$ in Fig. 6a).
    However, in most cases, the deviation is small.

The kinetic energy depends on both viscosity and magnetic energy.
The result shown in Fig. 4a that 
     the kinetic energy is almost independent of
     $\nu=\eta$ is actually the result of a rough cancellation between
     these two competing effects.
Suppose that magnetic field is turned off and, therefore,  the turbulence is
ordinary hydrodynamic turbulence. 
Let us also suppose that the viscosity is zero.
Then, at the energy injection scale, we will have
\be
 V^3  \sim \epsilon.
\ee
This is a well known relation in Kolmogorov phenomenology wherein
the energy injection rate $\epsilon$ is equal to 
the energy cascade rate at the largest energy containing scale,
which is proportional to the energy contained in the largest eddies ($\sim V^2$) 
divided by one eddy turnover time ($\sim L/V$).
When viscosity is not extremely small, we need to add a viscosity effect:
\be
 V^3 + \nu V^2 \sim \epsilon, 
\ee
where we omit constants of order unity in each term.
The second term on the left-hand side is proportional to dissipation at the
large scale.
Since the second term on the
left-hand side is much smaller than the first, we can assume that
$ \nu V^2 \propto \nu + O(\nu^2)$.
If we solve this relation for $V^2$, we obtain
\be
 V^2 \sim C_{\epsilon}\epsilon^{2/3} - C_{\nu} \nu, 
\ee
where the $C$'s are constants.
Now if we turn on the magnetic fields and set the viscosity to be zero, then
we have 
\be
 V^3 + B^2 V \sim \epsilon.
\ee
Here we also omit constants of order unity in each term.
The first term on the left-hand side is still proportional to $V^3$,
which is supported by the relation $T_{v|v_2} \propto V^3$
(Fig. 6a).
The second term on the left-hand side comes from the relation $T_{b|v_2} \propto VB^2$
(Fig. 6b).
Since the second term is small compared with the first,
we can write $ B^2 V \propto B^2 + O(B^4)$.
Solving for $V^2$, we have
\be
 V^2 \sim C_{\epsilon}\epsilon^{2/3} - C_{B} B^2. 
\ee
If we combine the two relations, we have
\be
 V^2 \sim C_{\epsilon}\epsilon^{2/3} - C_{\nu} \nu - C_{B} B^2,
\ee
where we ignore the interactions between the $\nu$-effect and the $B^2$-effect.
If any, such interactions will be weak because both the $\nu$-effect and the $B^2$-effect
are weak.
This equation implies that the quantity $V^2 + C_{B} B^2$ depends only on $\nu$.

To complete the above analysis, we need an expression for $B^2$.
To this end, let us adopt an energy cascade model analogous to the ordinary 
hydrodynamical case: The energy $transfer$ rate is equal to the $magnetic$ energy
cascade rate at the large scale. The energy transfer rate is given in
equation (\ref{stretching}). The magnetic energy cascade rate is $B^2/(L/V)$.
Therefore, we have
\be
  (V-B)B^2 \sim B^2V + \eta B^2,
\ee
where the second term on the right-hand side is the large scale 
dissipation term.  Again, we omit all order unity constants.
If we solve the equation for $B$, we obtain
\be
  B = C_V V - C_{\eta} \eta,
\ee
where the $C$'s are constants.
In fact, our calculation shows that
\be
 \frac{ B-0.6V }{ \eta } = -26.7 \pm 2.1,
\label{pred}
\ee
for Runs 128A, 96A, 72A, 64A, 64B, 64C, and 64D.
This implies that $C_V=0.6$ and $C_{\eta}\sim 26$.
This relation slightly underestimates the magnetic field
strength for Run 144A.  (It gives $B\sim 0.6V$, as opposed to
the observed value of $0.8V$.)  If we examine Fig. 7 we see
that this discrepancy arises from the difference between
the rms magnetic field strength, and the magnetic 
field strength on the largest eddy scale.  Comparing
the hyperviscosity run (144A) with the other cases, we see
that the former has a substantially greater fraction of
its magnetic energy in small scale structures.  If
we interpret $B$ in the preceding equation as the average
obtained by smoothing on the large eddy scale, which is
consistent with our derivation, then 
the hyperviscosity run is in agreement with equation (\ref{pred}).
In this sense, the values of $B$ and $B^2$ present in this
    discussion section should be regarded as lower limits for the
case of vanishing diffusivity. Alternatively, one might regard the
following discussion as valid for cases of moderate magnetic
    Reynolds numbers and that this analysis sets 
    limits for the case of zero diffusivity.
For $B^2$, we have
\be
  B^2 \sim C_V^2 V^2 - 2C_V C_{\eta}V\eta.
  \label{bsq}
\ee
Here we omitted $C_{\eta}^2 \eta^2$ on the right-hand side.

If we combine expressions for $V^2$ and $B^2$, we have
\begin{eqnarray}
  V^2 \sim 
    ( C_{\epsilon}\epsilon^{2/3} - C_{\nu} \nu + C_{\eta} \eta )/(1+C_B C_V^2), 
    \label{vsq}\\
  B^2 \sim 
    ( C_V^2 C_{\epsilon} \epsilon^{2/3} - C_V^2 C_{\nu} \nu - C_V C_{\eta}V\eta)
    /(1+C_B C_V^2).
\end{eqnarray}
Note that, even though we treat $\nu$ and $\eta$ separately, there is no guarantee that 
the above relations are true for non-unity Prandtl numbers.
Since $\nu=\eta$ and the second and third terms on the right-hand side of 
equation (\ref{vsq})
have different signs, $V^2$ depends on $\nu$ (=$\eta$) weakly.
We expect $C_B$ and $C_V$ have similar values because they are describing 
similar things - dissipation.
The expression for $B^2$ is not absolutely correct because the
omitted term in equation (\ref{bsq}) 
becomes non-negligible when $\eta$ becomes large.

When $\nu=\eta=0$, we have
\begin{eqnarray}
  V^2 \sim C_{\epsilon}\epsilon^{2/3}/ (1+C_B C_V^2),\\
  B^2 \sim C_V^2 C_{\epsilon} \epsilon^{2/3} / (1+C_B C_V^2).
\end{eqnarray}
These give a lower limit for $B^2$ and an upper limit for $V^2$.

\acknowledgements
This work was partially supported by National Computational Science
Alliance under CTS980010N and utilized the NCSA SGI/CRAY Origin2000.

\onecolumn
\clearpage

\begin{deluxetable}{llllclllllll}  
\footnotesize
\tablecaption{Results of Simulations}
\tablewidth{0pt}
\tablehead{
\colhead{Run} & \colhead{$N^3$} & \colhead{$\nu=\eta$} & 
\colhead{$B_0^2$} & \colhead{$V^2$} & \colhead{$B^2$} & 
\colhead{$\epsilon$} &
\colhead{$D_M$} & \colhead{$T_{v|v_2}$\tablenotemark{a}} & 
\colhead{$T_{b|v_2}$\tablenotemark{b}} &
\colhead{$k_e$\tablenotemark{c}} &
\colhead{($t_1,t_2$)\tablenotemark{d}}
}
\startdata
REF1&$96^3$&.0043& $10^{-4}$&.773 & .169& .167 &.088&.0815&.0353& -   &(500,560)\nl
REF2&$64^3$&.015 & -    &0.85 &- & .163 &-&-&-& -   &(300,750)\nl
REF3&$96^3$&.0043& -    &1.053 &- & .160 &-&-&-& -   &(50,165)\nl
144A&$144^3$&hyper& $10^{-7}$&.649& .420& .161 & - &.0519&.0758& 8.4 &(60,240)\nl
128A&$128^3$&.003 & $10^{-7}$&.755& .200& .166 &.099&.0794&.0421& 7.1&(100,210)\nl
96A &$96^3$&.0043& $10^{-7}$&.761& .170& .166 &.088&.0807&.0364& 6.7 &(200,500)\nl
72A &$72^3$&.0064& $10^{-7}$&.792& .122& .165 &.070&.0870&.0276& 6.8 &(300,800)\nl
64A &$64^3$&.0074& $10^{-7}$&.786& .113 & .166&.064&.0835&.0257& 6.6 &(300,800)\nl
64B &$64^3$&.01  & $10^{-7}$&.794& .0804& .165&.048&.0860&.0019& 6.6 &(300,750)\nl
64C &$64^3$&.015 & $10^{-7}$&.808& .0201& .164&.013&.0903&.0046& 8.0 &(300,800)\nl
64D &$64^3$&.02  & $10^{-7}$&.774&$<10^{-4}$& .164&$\sim10^{-5}$& -&-& - &(300,750)\nl
\enddata

\tablenotetext{a}{ $T_{v|v_2} \equiv T_{v|v_k}(k=2)$ }
\tablenotetext{b}{ $T_{b|v_2} \equiv T_{b|v_k}(k=2)$ }
\tablenotetext{c}{ Equipartition wavelength }
\tablenotetext{d}{Time interval used for averaging physical quantities}
             
\end{deluxetable}


\clearpage
\figcaption{Test of the code. Time evolution of $T_{b|v}$, $T_{v|b}$, 
            and magnetic dissipation.
            $T_{b|v}$ and $-T_{v|b}$ coincide exactly. Magnetic dissipation shows
            a time delay, which is the turbulent diffusion time scale.
            Run REF1.}
\figcaption{Time evolution of kinetic and magnetic energy. The results
 at three different resolutions are shown: dotted, $144^3$ (Run 144A);
 solid, $96^3$ (Run 96A); dashed, $64^3$ (Run 64A).
 For 64A and 96A, the kinetic energy has very similar values. 
 The magnetic energy depends on $\nu$ $(=\eta)$. In the case of hyperviscosity 
 (Run 144A), the magnetic energy is more than half of the kinetic energy.}
 
\figcaption{Time evolution of $kT_{b|v_k}(k)$ and $E_M(k)$. Run 144A}
 
\figcaption{Dependence of physical quantities on $\nu$ $(=\eta)$.
 $Top$: The kinetic energy density is nearly independent of  the physical viscosity.
 The magnetic energy density grows as $\nu$ $(=\eta)$ decreases.
 $Bottom$: This relation implies $D_M \sim (V-B/1.6)B^2$.}
 
\figcaption{Energy transfer and flux spectra for Run 144A. 
 $Top$: Energy transfer spectra. The spectrum of $T_{b|v_k}$ (dotted line) shows that
        magnetic fields gain energy from large scale eddies.
 $bottom$: Energy flux spectra. The spectrum of $\Pi_{b|v_k}$ (dotted line) rises 
        rapidly at small $k$ and reaches the value of $\epsilon$ at $k\sim 10$.}
 
\figcaption{Evidences that large-scale turbulence is similar to ordinary 
  hydrodynamic turbulence. 
 $Top$: $T_{v|v_k}(k=2)/V^3$ is nearly constant for
 nonzero physical viscosity, which implies that $T_{v|v_k}(k=2)\sim V^3$.
 In the case of hyperviscosity, this relation may not hold true.
 $Bottom$: $T_{b|v_k}(k=2)/(B^2V)$. This relation implies stretching of
 magnetic field lines without significant back reaction 
 is responsible for the nonlinear energy transfer from
 velocity components at $k\sim 2$ to magnetic fields.} 
 
\figcaption{Energy spectra.  The results
 at three different resolutions are shown: dotted, $144^3$ (Run 144A);
 solid, $96^3$ (Run 96A); dashed, $64^3$ (Run 64A). Both axes
 are drawn in logarithmic scale. }

\figcaption{Second order structure functions for the magnetic and
velocity fields.  The horizontal axis is in the direction of the
local magnetic field, defined by pairwise averaging.  The velocity
fields show correlations at all scales, with a smooth transition
from isotropy on large scales to anisotropy on small scales.  The
magnetic field shows similar behavior on small scales, below
the equipartition scale, but has almost no correlation on large
scales.}

\clearpage
\begin{figure}
\plotone{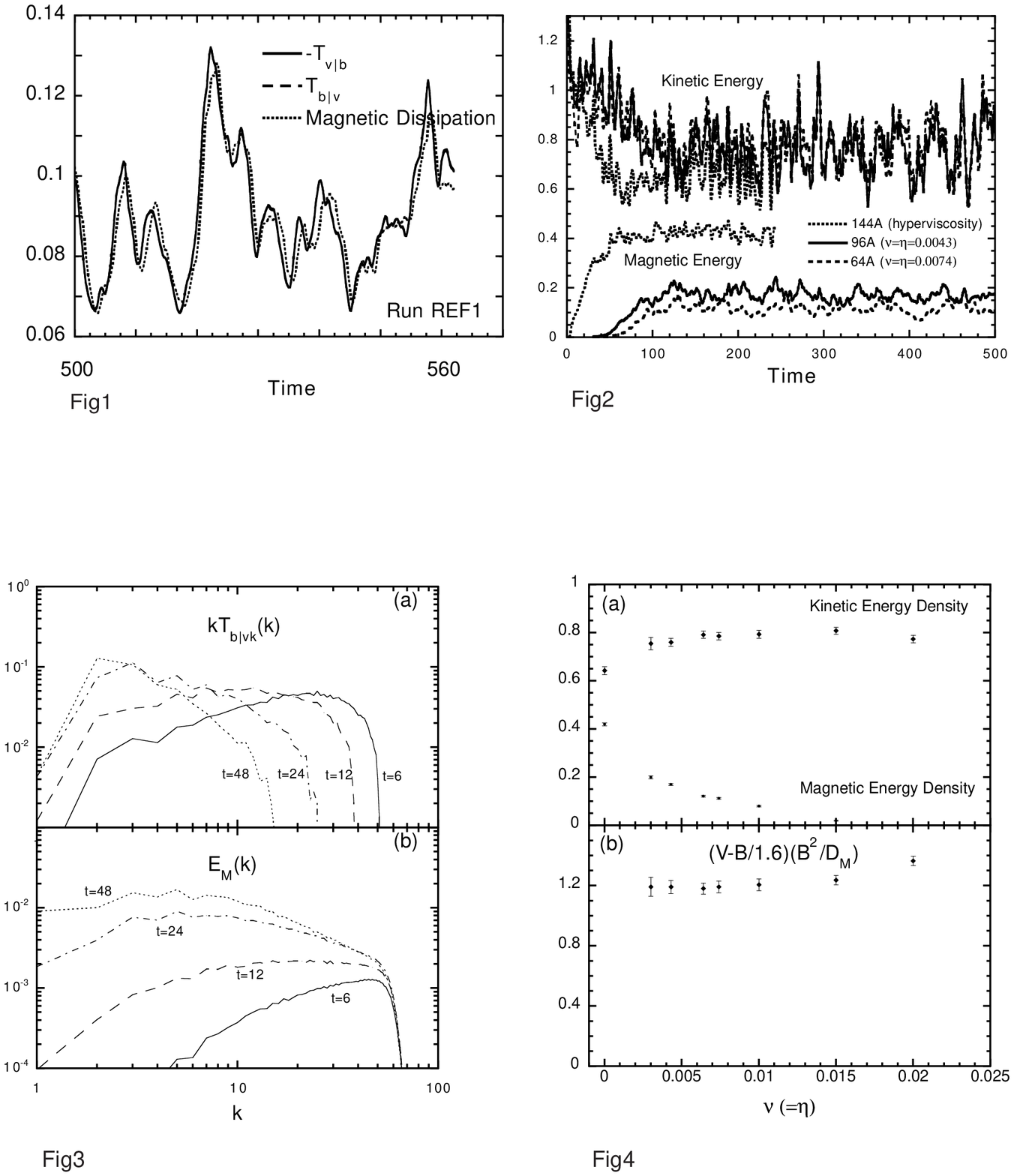}
\end{figure}
\clearpage
\begin{figure}
\plotone{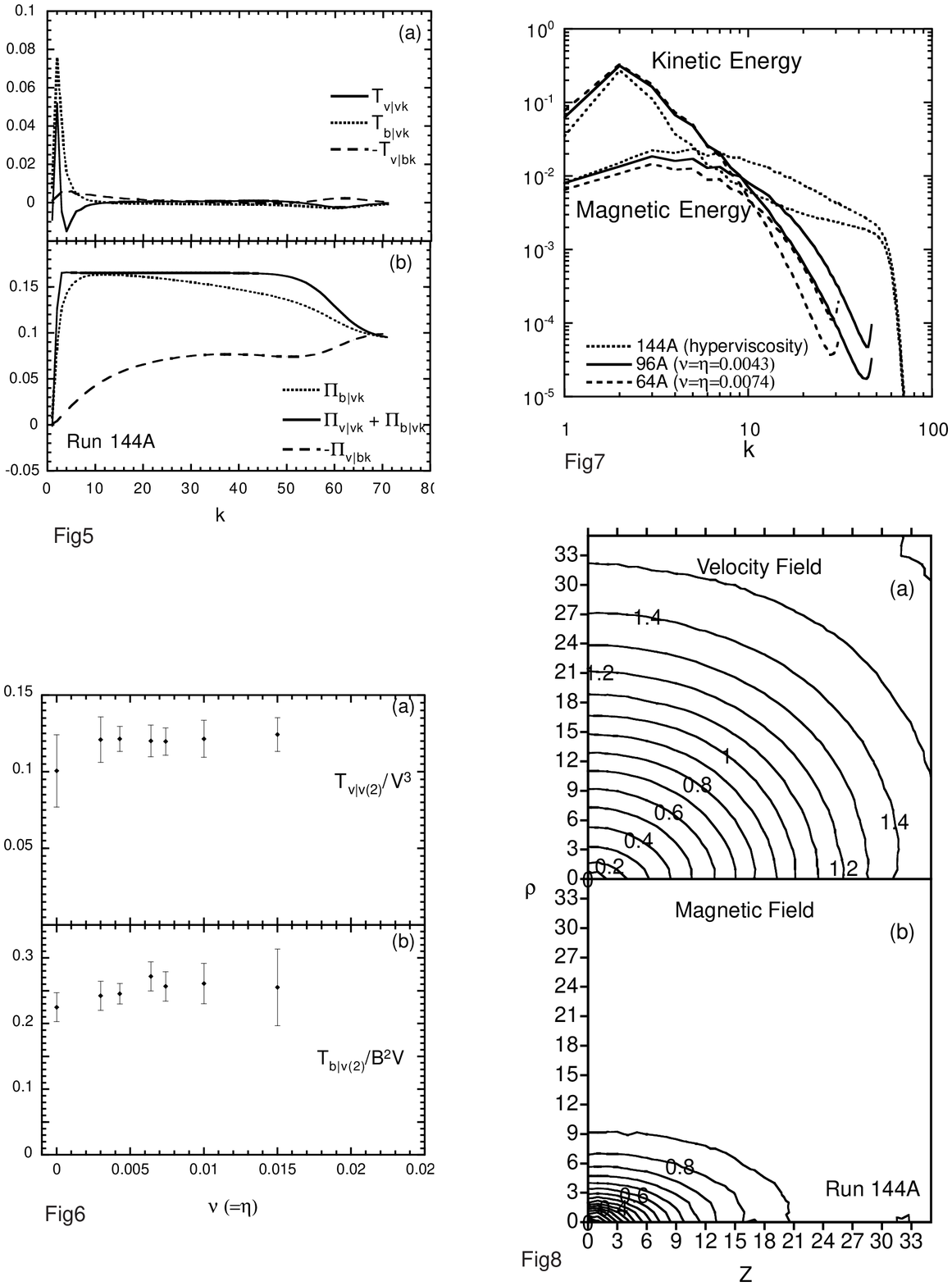}
\end{figure}

\end{document}